\documentclass[twocolumn,aps,prb,showpacs]{revtex4}

\usepackage{graphicx}
\usepackage{amsmath,amssymb}
\usepackage{hyperref}
\usepackage{setspace}

\begin{document}

\title{Temperature Dependence of Interlayer Magnetoresistance in Anisotropic Layered Metals}
\author{Braden A. W. Brinkman and Malcolm P. Kennett}
\affiliation{Department of Physics, Simon Fraser University,\\8888 University Drive, Burnaby, British Columbia, V5A 1S6}
\date{\today}

\begin{abstract}
Studies of interlayer transport in layered metals have 
generally made use of zero temperature conductivity expressions to analyze
 angle-dependent magnetoresistance oscillations (AMRO).  However, recent 
high temperature AMRO experiments have been performed in a regime where 
the inclusion of
finite temperature effects may be required for a quantitative 
description of the resistivity.
We calculate the interlayer conductivity in a layered metal with 
anisotropic Fermi surface properties allowing for finite temperature 
effects.  We find that resistance maxima are modified by thermal effects
much more strongly than resistance minima.
 We also use our expressions to calculate the interlayer resistivity 
appropriate to recent AMRO experiments in an overdoped cuprate which 
led to the conclusion that there is an anisotropic, linear in temperature 
contribution to the scattering rate and find that this conclusion is 
robust.
\end{abstract}

\pacs{71.18.+y, 72.10.-d, 72.15.-v, 74.72.-h}

\maketitle

\section{Introduction}
\label{sec:intro} 

Angle-dependent magnetoresistance oscillations (AMRO) in the interlayer resistance
 of layered metals provide a means to determine Fermi surface properties of such 
systems.\cite{kart}  Maps of two and three dimensional Fermi surfaces have been 
determined for a wide variety of materials.\cite{berg,balicas,osada,beierl,husseynature,enomoto} 
 AMRO can also be used to infer information about possible anisotropies in the 
momentum space of Fermi surface properties such as the scattering 
rate.\cite{Sandemann,ajawad,kennettmckenzie,ajawad2,analytisetal,kennett2,hussey2,singleton,smithmckenzie} 
 In particular, recent AMRO experiments by Abdel-Jawad \emph{et al.} \cite{ajawad} 
reveal that in the overdoped cuprate Tl$_2$Ba$_2$CuO$_{6+\delta}$ (Tl2201) the 
scattering rate contains an anisotropic piece that varies with the same symmetry 
as the superconducting gap and grows linearly with temperature.  The 
measured anisotropy in the scattering rate may have relevance to the 
origin of high temperature superconductivity in cuprate superconductors, 
as the strength of the anisotropic 
piece of the scattering increases with the critical temperature, $T_c$.\cite{ajawad2}  
Scattering rate anisotropy has been detected in several other cuprates 
using angle-resolved photoemission spectroscopy (ARPES),\cite{plate,kaminski,kondo,Chang} 
optical conductivity\cite{hac} and in-plane transport measurements.\cite{narduzzo}  
Theoretical support for such anisotropic scattering comes from ideas about 
hot and cold spots on the Fermi surface, first raised in the context of optical 
conductivity,\cite{ioffe} and appears to arise naturally in dynamical mean field 
calculations on the two dimensional Hubbard model.\cite{civelli}

An advantage of AMRO over other techniques which use oscillations to obtain information
about the Fermi surface, such as de Haas-van Alphen and Shubnikov-de Haas oscillations, 
is that they can be measured at elevated temperatures, provided $\omega_c\tau$, the
combination of the cyclotron frequency and the transport lifetime is large enough.
To date, expressions used to fit AMRO data have been derived assuming zero temperature, which 
should work reasonably well for temperatures $T \ll T_F$, the Fermi temperature.
Recently there have been relatively high temperature AMRO experiments for which 
corrections to the zero temperature approximation may be needed. For instance, 
the experiments in Refs.~\onlinecite{ajawad,analytisetal} 
were performed for $T$ up to 55 K,
corresponding to a maximum thermal energy of approximately 0.02 of
the Fermi energy $\varepsilon_F = k_B T_F$.  At this temperature scale, one expects  
the zero-temperature approximation to work reasonably well in fits to AMRO.
However, more recent experiments\cite{private}
have extended AMRO to temperatures as high as 110 K ($\sim 0.04 \, T/T_F$)
where finite temperature effects should be more relevant --
we show below that finite temperature corrections to AMRO can be quantitatively
important at these temperatures.

In this paper
we generalize existing zero temperature semi-classical AMRO formulae for
layered metals with anisotropic Fermi surface properties to finite temperatures.
The finite temperature expression for interlayer conductivity with anisotropic 
Fermi surface properties is our main result.  We use this expression to generate 
numerically the AMRO expected in overdoped thallium cuprate for an isotropic 
scattering rate.  Fitting to the simulated AMRO with zero temperature expressions 
reveals no significant anisotropic contribution to the scattering rate.  Our results
indicate that the inferred scattering rates in Ref.~\onlinecite{ajawad} are robust
to finite temperature corrections, but that finite temperature corrections will be
required in fits to AMRO at higher temperatures.

This paper is structured as follows: in Sec.~\ref{sec:CalculationOfAMRO} we present 
our analytic calculations of AMRO.  In Sec.~\ref{sec:numerics} we perform 
numerical calculations to check whether the finite temperature effects can 
masquerade as linear in $T$ anisotropic scattering.  In Sec.~\ref{sec:conclusion} 
we summarize our results and conclude.

\section{Calculation of AMRO}
\label{sec:CalculationOfAMRO}
In this section we briefly review the calculation of the interlayer conductivity in a 
layered metal allowing for non-zero temperature and treat the cases of isotropic
Fermi surface properties (for which we can make considerable analytic progress) and anisotropic
Fermi surface properties separately.
We assume a simple model of a layered metal in which the $c$-axis is the weakly conducting direction and take a dispersion relation of the form
\begin{equation}
\varepsilon(\mathbf{k}) = \varepsilon_{2d}(k_x,k_y) - 2t_c(k_x,k_y)\cos(ck_z),
\label{eqn:3ddisp}
\end{equation}
where $\mathbf{k} = (k_x,k_y,k_z)$ is the electron wavevector, $\varepsilon_{2d}$ is the dispersion in the $k_x$-$k_y$ plane, $t_c(k_x,k_y)$ is the (possibly anisotropic) interlayer hopping term (assumed to be small compared to the Fermi energy), and the parameter $c$ is the distance between the conducting layers.

\begin{figure}[htb] 
\centering
  \includegraphics[scale=0.8]{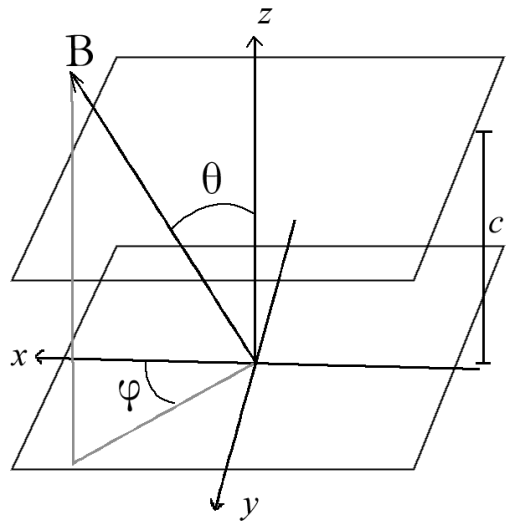}
  \caption{Schematic of the magnetic field in an AMRO experiment.  
The polar angle $\theta$ is measured relative to the $z$-axis and 
the azimuthal angle $\varphi$ relative to the $x$-axis, which lies in the conducting plane.}
     \label{fig:metaldiag}
\end{figure}

We treat the electrons semiclassically and calculate AMRO by solving the time independent and 
spatially uniform Boltzmann Equation in the relaxation time approximation:
\begin{equation}
\mathbf{F} \cdot \frac{\partial f}{\partial \mathbf{p}} = -\frac{f - f_T}{\tau},
\label{eqn:boltzmann}
\end{equation}
where $f(\mathbf{p})$ is the electron distribution function, in terms of 
momentum $\mathbf{p}$,  $f_T$ is the Fermi-Dirac distribution, 
$\mathbf{F} = -e(E_z\mathbf{\hat{z}} + \mathbf{v} \times \mathbf{B})$ 
is the Lorentz force for a magnetic field $\mathbf{B}$ and a weak electric 
field perpendicular to the conducting layers, $E_z$, 
$\mathbf{v} = \hbar^{-1}\nabla_{\mathbf{k}}\varepsilon(\mathbf{k})$ is the velocity, 
and $\tau$ is the transport lifetime.  In general $\tau$ may depend on both 
the electron momentum and energy.  We assume that the magnetic field 
$\mathbf{B} = B\left( \sin \theta \cos \varphi, \sin \theta \sin \varphi, 
\cos \theta \right)$ is applied at an angle $\theta$ with respect to the 
$z$-axis and an angle $\varphi$ with respect to the $x$ axis, as shown
schematically in Fig.~\ref{fig:metaldiag}.

For weak electric fields we can write $f \approx f_T + \delta f$, where 
$\delta f \sim \mathcal{O}(E_z)$.  Neglecting small terms in the 
electric field and the hopping, $t_c$, we obtain
\begin{equation}
\frac{\partial f}{\partial \phi} + \frac{f}{\omega_0(\varepsilon,\phi)\tau(\varepsilon,\phi)} 
= -\frac{ev_zE_z}{\omega_0(\varepsilon,\phi)}
\left(-\frac{\partial f_T}{\partial \varepsilon}\right),
\label{eqn:generalboltz}
\end{equation}
where $\phi$ is the angular position in momentum space and $\varepsilon$ is the energy.  
We relabel $\delta f \rightarrow f$ for convenience, and define\cite{kennettmckenzie}
$$\omega_0(\varepsilon,\phi) = eB\cos\theta\frac{\mathbf{k}_{\parallel}(\varepsilon,\phi)
\cdot\mathbf{v}_{\parallel}(\varepsilon,\phi)}{\hbar|\mathbf{k}_{\parallel}(\varepsilon,\phi)|^2},$$
with $\mathbf{v}_\parallel = (v_x,v_y)$ and $\mathbf{k}_\parallel = (k_x,k_y)$.  The in-plane 
dispersion relation $\varepsilon_{2d}(k_x,k_y)$ determines $\mathbf{v}_\parallel$ and 
$\mathbf{k}_\parallel$; accordingly, if $\varepsilon_{2d}$ is anisotropic, 
$\mathbf{v}_\parallel$, $\mathbf{k}_\parallel$ and $\omega_0(\varepsilon,\phi)$ will also be 
anisotropic in momentum space.

Equation~(\ref{eqn:generalboltz}) holds for both isotropic and anisotropic Fermi surfaces. 
 The formal solution to Eq.~(\ref{eqn:generalboltz}) is
\begin{eqnarray}
f(\phi,\varepsilon) & = & -eE_z\int_0^\infty d\epsilon 
\int_{-\infty}^{\phi}d\phi'~G(\phi,\phi') 
\frac{v_z(\varepsilon,\phi')}{\omega_0(\varepsilon,\phi')}
\left(-\frac{\partial f_T}{\partial \varepsilon}\right),\nonumber \\
\label{eqn:dist}
\end{eqnarray}
where 
\begin{equation}
G(\phi,\phi',\varepsilon) = \exp\left[{-\int_{\phi'}^{\phi}\frac{d\psi}{\omega_0(\varepsilon,\psi)
\tau(\varepsilon,\psi)}}\right],
\label{eqn:green}
\end{equation}
is the probability that an electron will travel around the Fermi surface from an angle 
$\phi'$ to $\phi$ without being scattered.\cite{kennettmckenzie}
With this expression we may calculate the current density 
$j_z = -2e\int\frac{d^3\mathbf{k}}{(2\pi)^3}v_zf$ and hence the interlayer 
conductivity $\sigma_{zz}$.  In Sec.~\ref{sec:isofermi} and 
Sec.~\ref{sec:anisofermi} we present the resulting finite temperature interlayer 
conductivity for both isotropic and anisotropic cases, respectively.

\subsection{Isotropic Fermi Surface}
\label{sec:isofermi}
We first consider finite temperature effects for an isotropic layered metal, with 
isotropic scattering rate $1/\tau$ (we allow $\tau$ to depend on $\varepsilon$), isotropic 
hopping $t_c$ and an isotropic dispersion,\cite{footnote}
\begin{equation}
\varepsilon_{2d}(k_x,k_y) = \frac{\hbar^2}{2 m^\ast}\left(k_x^2+k_y^2\right),
\label{eqn:2ddisp}
\end{equation}
where $m^\ast$ is the effective mass.  This gives $\omega_0 = eB\cos\theta/m^\ast$.

In the zero temperature limit the expression for the interlayer
 conductivity is\cite{mckenziemoses,yagi}
\begin{equation}
\sigma_{zz}(\theta) = \sigma_0\left[J_0^2(\gamma k_F)+ 2\sum_{s=1}^{\infty}
\frac{J_s^2(\gamma k_F)}{1 + (s\omega_0\tau)^2}\right],
\label{eqn:cond}
\end{equation}
where $\sigma_0 = \frac{2e^2m^\ast c t_c^2 \tau}{\pi \hbar^4}$ ($\tau$, if energy 
dependent, is evaluated at the Fermi energy) and $\gamma = c\tan\theta$. 
 To generalize this expression to finite temperature we must integrate over energy in 
calculating $\sigma_{zz}$.  The integral to be evaluated is
\begin{eqnarray}
\sigma_{zz} & = & \vartheta_0\int_0^{\infty}d\varepsilon 
\left(-\frac{\partial f_T}{\partial \varepsilon}\right)\tau(\varepsilon) \nonumber \\
& & \times \left[J_0^2(\lambda \sqrt{\varepsilon})+ 2\sum_{s=1}^{\infty}\frac{J_s^2(\lambda 
\sqrt{\varepsilon})}{1 + (s\omega_0\tau(\varepsilon))^2}\right], \nonumber \\
\label{eqn:isoenergy}
\end{eqnarray}
where the energy dependence of the scattering rate $\tau^{-1}$ must be taken into 
account, $\vartheta_0 =2e^2m^\ast c t_c^2/\pi \hbar^4$, and $\lambda = 
\frac{\gamma k_F}{\sqrt{\varepsilon_F}}$.  Using a Sommerfeld expansion\cite{ashcroftmermin} 
and keeping terms to order $(T/T_F)^2$, we arrive at the result
\begin{eqnarray}
\sigma_{zz}  & \simeq &  \vartheta_0\left\{\tau(\mu)\left[J_0^2(\lambda\sqrt{\mu})+ 
2\sum_{s=1}^{\infty}\frac{J_s^2(\lambda \sqrt{\mu})}{1 + (s\omega_0\tau(\mu))^2}\right]\right. 
\nonumber \\ 
& & \hspace*{0.5cm}+~\frac{\pi^2}{6}(k_BT)^2\left.\left[\Psi_0(\varepsilon_F) + 
2\sum_{s=1}^{\infty}\Psi_s(\varepsilon_F)\right]\right\}, \nonumber \\
\label{eqn:result}
\end{eqnarray}
where the chemical potential is $$\mu = \varepsilon_F\left(1-
\frac{\pi^2}{12}\left(\frac{T}{T_F}\right)^2 + \ldots \right),$$ and we introduce
\begin{equation}
\Psi_s(\varepsilon) = \frac{d^2}{d\varepsilon^2}\left[\tau(\varepsilon)
\frac{J_s^2(\lambda \sqrt{\varepsilon})}{1+(s\omega_0\tau(\varepsilon))^2}\right].
\label{eqn:psis}
\end{equation}

\begin{figure}[htb]
\begin{center}
  \includegraphics[scale=0.9]{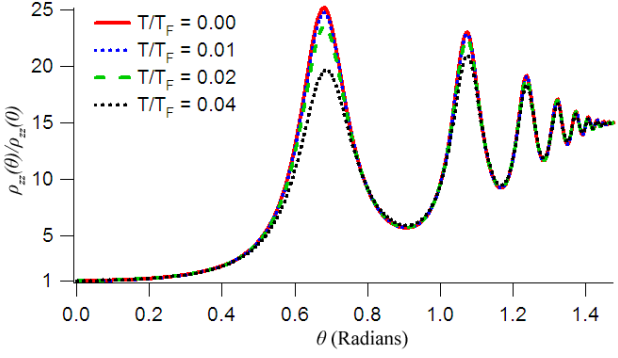}
\end{center}
  \caption{Plot of $\rho_{zz}/\rho_{0}$ versus $\theta$ for an
isotropic Fermi surface with $T/T_F=0,0.01,0.02$ and $0.04$.}
     \label{fig:resvstheta}
\end{figure}

In Fig.~\ref{fig:resvstheta} we plot the interlayer 
resistivity $\rho_{zz} = 1/\sigma_{zz}$ (normalized by $\rho_0 = 1/\sigma_0$), 
determined from Eqs.~(\ref{eqn:cond}) and (\ref{eqn:result}) for an energy-independent 
$\tau$ at several different temperatures.  We choose $c k_F = 3$ and
$\omega_c \tau = 5$ (where $\omega_c = eB/m^\ast$) to match values used in 
plots in Ref.~\onlinecite{mosesandmckenzie} and 
 plot temperatures of 0, $0.01\hspace*{0.1cm}T_F$, $0.02\hspace*{0.1cm}T_F$ 
and $0.04\hspace*{0.1cm}T_F$.  Deviations from the zero temperature result are 
mainly noticeable in the first few AMRO peaks, for $T/T_F \gtrsim 0.02$.  
Interestingly, the minima seem much less affected by finite temperature than the maxima.

To see why the maxima are affected more than minima we perform an asymptotic expansion 
of Eq.~(\ref{eqn:result}) in the limit $ck_F\tan\theta \rightarrow \infty$ 
for an energy independent $\tau$.  The derivation follows that for 
the $T=0$ case in Ref.~\onlinecite{mosesandmckenzie}.  We find the 
correction to the zero temperature result to be 
(when $(ck_F\tan\theta)\left(\frac{T}{T_F}\right)^2 \ll 1$)

\begin{eqnarray}
\Delta \sigma_{zz} & = & \frac{\vartheta_0}{ck_F \tan\theta} \frac{\pi^2}{6}
\left(\frac{T}{T_F}\right)^2  \frac{1}{\sinh\left(\frac{\pi}{\omega_c\tau\cos\theta}\right)} 
\nonumber \\
& &  \times
\Bigg[\cosh\left(\frac{\pi}{\omega_c\tau\cos\theta}\right)  \nonumber \\
& &  \hspace*{0.3cm}
 + \left(1 - (ck_F\tan\theta)^2\right)\sin\left(2ck_F \tan\theta\right) \nonumber \\
& &  \hspace*{1cm}
  - 2ck_F \tan\theta \cos(2ck_F\tan\theta)
\Bigg].
\label{eqn:asymcor}
\end{eqnarray}
The extrema of the resistivity occur at the 
Yamaji angles,\cite{Yamaji} which satisfy
\begin{equation}
ck_F\tan\theta_n= \pi\left(n + \frac{\nu}{4}\right),
\label{eqn:yamajiangle}
\end{equation}
where $\nu = 1~(3)$ corresponds to a minimum (maximum) of 
the resisitivity $\rho_{zz} = 1/\sigma_{zz}$.  
[Note that there will be ${\mathcal O}(T^2/T_F^2)$ corrections to the 
Yamaji angles, although as can be seen in Fig.~\ref{fig:resvstheta}, these
are not large, and they are not important for our discussion here.]
The convention used here for $n$ is that the $n^{\rm th}$ minimum 
follows the $n^{\rm th}$ maximum and $n$ begins at zero.  
We can calculate the $\mathcal{O}(T^2/T_F^2)$ correction to the extrema 
of the resistivity using Eq.~(\ref{eqn:asymcor}).  We find that at maxima the
relative change in the resistance is
\begin{widetext}
\begin{eqnarray}
\frac{\Delta\rho_{zz}^{\rm max}(T\neq 0)}{\rho_{zz}^{\rm max}(T=0)} & = &
-\frac{\pi^2}{6} \left(\frac{T}{T_F}\right)^2 
\frac{1}{\cosh\left(\frac{\pi}{\omega_c\tau\cos\theta_n^{\rm max}}
\right) - 1} 
\left[\cosh\left(\frac{\pi}{\omega_c\tau\cos\theta_n^{\rm max}}\right) + 
\pi^2 \left(n + \frac{3}{4}\right)^2\right] ,
\end{eqnarray}
and at minima the relative change in resistance is
\begin{eqnarray}
\frac{\Delta\rho_{zz}^{\rm min}(T\neq 0)}{\rho_{zz}^{\rm min}(T=0)} & = &
-\frac{\pi^2}{6} \left(\frac{T}{T_F}\right)^2 
\frac{1}{\cosh\left(
\frac{\pi}{\omega_c\tau\cos\theta_n^{\rm min}} \right) + 1} 
\left[\cosh\left(
\frac{\pi}{\omega_c\tau\cos\theta_n^{\rm min}}\right) - 
\pi^2 \left(n + \frac{1}{4}\right)^2\right]  .
\end{eqnarray}
\end{widetext}

In general the numerators in the two expressions will be comparable, but the denominator
in the minima expression is always greater than 2, whereas the denominator in the maxima 
expression can become very small when $\omega_c\tau\cos\theta_n^{\rm max}$ becomes large,
leading to a much stronger reduction in the resistance at the maxima than increase at the
resistance minima.  [It should be noted that assuming temperature independent scattering,
as we do in this example, is somewhat artificial (in general one 
expects the scattering rate to have a $T^2$ contribution in a Fermi liquid), but it
serves to illustrate the differing effects of temperature on resistance minima
and maxima.]

\subsection{Anisotropic Fermi Surface}
\label{sec:anisofermi}
We now turn to consider finite temperature effects for an anisotropic Fermi surface.  
We allow $\omega_0$, $\tau$, $\mathbf{v}_\parallel$, $\mathbf{k}_\parallel$ 
and $t_c$ to vary with $\phi$ and $\varepsilon$.  The zero temperature interlayer
conductivity for a layered metal with anisotropic Fermi surface properties was 
calculated by Kennett and McKenzie \cite{kennettmckenzie} to be
\begin{eqnarray}
\sigma_{zz} & = & \frac{s_0 eB\cos\theta}{1-P}\int_0^{2\pi}d\phi
\int_{\phi-2\pi}^{\phi}d\phi' \nonumber \\
& & \times \frac{t_c(\phi)t_c(\phi')}{\omega_0(\phi)
\omega_0(\phi')}G(\phi,\phi',\varepsilon_F)\cos\Phi(\phi,\phi'), \nonumber \\
\label{eqn:result1}
\end{eqnarray}
where $s_0 = e^2c/(\pi\hbar^4)$, $P = G(2\pi,0,\varepsilon_F)$ 
is the probability that an electron will make
 a full cyclotron orbit about the Fermi surface before being scattered and 
\begin{eqnarray*}
\Phi(\phi,\phi') & = & \gamma \mathbf{\hat{b}}_{\parallel}(\varphi)
\cdot\left(\mathbf{k}_F(\phi)-\mathbf{k}_F(\phi')\right) \\
& = &  \gamma (k_F(\phi)\cos(\varphi-\phi) - k_F(\phi')\cos(\varphi-\phi')),
\end{eqnarray*} 
where we have defined $\hat{\mathbf{b}}_\parallel(\varphi) = (\cos \varphi, \sin \varphi)$
as the direction of the component of the magnetic field in the $x$-$y$ plane, and 
$\mathbf{k}_F(\phi) = k_F(\phi)(\cos\phi,\sin\phi)$, where $k_F(\phi)$ is 
the magnitude of the Fermi wavevector.

Generalizing Eq.~(\ref{eqn:result1}) for the interlayer conductivity to finite temperatures 
is similar to the isotropic case, and again we integrate over energy:
\begin{eqnarray}
\sigma_{zz} & = & s_0 eB\cos\theta\int_{-\infty}^{\infty}d\varepsilon\int_0^{2\pi}
d\phi\int_{\phi-2\pi}^{\phi}d\phi'~\left(-\frac{\partial f_T}{\partial \varepsilon}\right) \nonumber
\\ & & \times
\frac{t_c(\varepsilon,\phi)t_c(\varepsilon,\phi')}{\omega_0(\varepsilon,\phi)
\omega_0(\varepsilon,\phi')}\frac{G(\phi,\phi',\varepsilon)}{1-P(\varepsilon)}
\cos\Phi(\phi,\phi',\varepsilon). \nonumber \\ & &
\label{eqn:result2}
\end{eqnarray}
The factor $G(\phi,\phi',\varepsilon)$ is as in Eq.~(\ref{eqn:green}), and the 
energy-dependent $\Phi(\phi,\phi',\varepsilon)$ is
\begin{eqnarray}
\Phi(\phi,\phi',\varepsilon) & = & \gamma (\kappa(\varepsilon,\phi)\cos(\varphi-\phi) \nonumber \\
& & \hspace*{0.5cm}-~\kappa(\varepsilon,\phi')\cos(\varphi-\phi')),
\label{eqn:phase}
\end{eqnarray}
where $\kappa(\varepsilon,\phi) \equiv |\mathbf{k}_\parallel| = \sqrt{k_x^2 + k_y^2}$ is the magnitude of the electron wavevector.  Allowed values of $\kappa$ are determined from the dispersion relation.  Equation~(\ref{eqn:result2})
 is our main result and is independent of whether the interlayer hopping is coherent 
(assumed here) or weakly incoherent.\cite{kennettmckenzie,mckenziemoses,mosesandmckenzie}  
The only difference in AMRO for the two models of transport occurs for
angles near $\theta = \frac{\pi}{2}$.\cite{kart,mckenziemoses,mosesandmckenzie,kennett4}

In order to make analytic progress with Eq.~(\ref{eqn:result2}) we need to specify the energy and
 $\phi$ dependence of various Fermi surface properties. In general, 
AMRO must be calculated numerically, although under certain conditions we may derive 
asymptotic expressions.  To do so, we note that if $\omega_0$ never approaches zero, 
the largest contribution to the integrand of Eq.~(\ref{eqn:result2}) will be 
from near $\varepsilon = \mu$, where the derivative of the Fermi-Dirac function 
is sharply peaked.  We can then perform a Sommerfeld expansion about this point, 
as in the isotropic finite temperature case.  The zeroth order term gives 
the zero temperature result evaluated at $\mu$ instead of $\varepsilon_F$.  
Under the conditions
$$c\kappa(\mu,\phi)\tan\theta \gg 1, \quad 
\frac{eBc|\mathbf{v}_\parallel(\mu,\phi)|\tau(\mu,\phi)}{\hbar} \gg 1,$$ 
 we can evaluate the zeroth 
order term using a stationary phase calculation.\cite{kennettmckenzie}  
To second order in $T/T_F$ we can then write the temperature 
dependent conductivity, valid near 
$\theta = \frac{\pi}{2}$, for a Fermi surface with $\phi \rightarrow \phi + \frac{\pi}{2}$ 
symmetry as 
\begin{widetext}
\begin{eqnarray}
\sigma_{zz}(\theta,\varphi,T) & \simeq & \frac{eBs_0\cos\theta}{1-P(\mu)}\frac{2\pi|\eta(\mu,\phi_0)|^{-1}}{c\tan\theta}\left(\frac{t_c(\mu,\phi_0)}{\omega_0(\mu,\phi_0)}\right)^2 \left[1+P(\mu)+2\sqrt{P(\mu)}\sin\left(2c\tan\theta \kappa(\mu,\phi_0)\cos(\varphi-\phi_0)\right)\right]\nonumber \\    
& & +~\frac{\pi^2}{6}eBs_0\cos\theta(k_BT)^2\Omega(\varepsilon_F),
\label{eqn:anisotemp}
\end{eqnarray}
where $\phi_0$ satisfies $\frac{\partial}{\partial \phi}\left[\mathbf{\hat{b}}_{\parallel}(\varphi)\cdot\mathbf{k}_\parallel(\varepsilon,\phi)\right] = 0$, $\eta(\varepsilon,\phi) \equiv \frac{\partial^2}{\partial \phi^2}\left[\mathbf{\hat{b}}_{\parallel}(\varphi)\cdot\mathbf{k}_\parallel(\phi)\right]$, and
\begin{eqnarray}
\Omega(\varepsilon) & = & \frac{d^2}{d\varepsilon^2}\left[\int_0^{2\pi}d\phi\int_{\phi-2\pi}^{\phi}d\phi'~\frac{t_c(\varepsilon,\phi)t_c(\varepsilon,\phi')}{\omega_0(\varepsilon,\phi)\omega_0(\varepsilon,\phi')}\frac{G(\phi,\phi',\varepsilon)}{1-P(\varepsilon)}\cos\Phi(\phi,\phi',\varepsilon)\right] \nonumber \\
 & \simeq & \frac{eBs_0\cos\theta}{c\tan\theta}\frac{d^2}{d\varepsilon^2}\left\{\frac{2\pi|\eta(\varepsilon,\phi_0)|^{-1}}{1-P(\varepsilon) }\left(\frac{t_c(\varepsilon,\phi_0)}{\omega_0(\varepsilon,\phi_0)}\right)^2 \left[1+P(\varepsilon)+2\sqrt{P(\varepsilon)}\sin\left(2c\tan\theta \kappa(\varepsilon,\phi_0)\cos(\varphi-\phi_0)\right)\right]\right\}. \nonumber \\
\label{eqn:correction}
\end{eqnarray}
\end{widetext}

\section{Numerics}
\label{sec:numerics}
We now outline the numerical scheme we use to test whether the temperature 
dependence of the conductivity affects the anisotropy obtained from fits to 
the scattering rate $1/\tau$ using zero temperature expressions for the 
conductivity.  Our approach is as follows: we use the tight-binding 
dispersion relation inferred from ARPES\cite{plate} to simulate the 
AMRO data we expect to observe at several different temperatures and 
azimuthal field angles $\varphi$, assuming an isotropic scattering rate 
of the form predicted by Fermi liquid theory.  We then fit to this 
data using the same procedure used to fit experimental data in Ref.~\onlinecite{ajawad}. 
First, we allow all parameters to vary in fitting the lowest temperature data 
(using the zero temperature expression for the interlayer resistivity).  
Second, the higher temperature data is fitted assuming only the scattering 
rate is temperature dependent.  If the scattering rate anisotropy 
is a fitting artifact, fits to AMRO data simulated with finite 
temperature expressions for the conductivity should yield 
results similar to the fits to the real data: an anisotropic 
contribution to the scattering rate that varies as $T\cos 4\phi$. 

The tight binding dispersion relation determined by Plat\'{e} \emph{et al}.\cite{plate} is:
\begin{eqnarray}
\varepsilon_{2d}(k_x,k_y) & = &  \mu + \frac{t_1}{2}(\cos k_x + \cos k_y) + t_2(\cos k_x \cos k_y)
\nonumber \\ &  &  + \frac{t_3}{2}(\cos 2 k_x + \cos 2 k_y)\nonumber \\ 
& &  + \frac{t_4}{2}(\cos 2k_x \cos k_y + \cos k_x \cos 2k_y \nonumber \\
& & \hspace*{2cm} + \cos k_x \cos k_y) \nonumber \\
& & + t_5 \cos 2k_x \cos 2k_y,
\label{eqn:tightbinding}
\end{eqnarray}
where the wavenumbers are measured in units of the in-plane lattice constant $a$, the $t_i$ are hopping parameters and $\mu$ is the chemical potential.  The values of the parameters are $\mu = 0.2438$, $t_1 =-0.725$, $t_2 =0.302$, $t_3=0.0159$, $t_4=-0.0805$ and $t_5=0.0034$, all in eV.  We use $\varepsilon_{2d}$ to calculate $1/\omega_0$ and $\kappa$ as functions of $\phi$ numerically, for several energies.  We fit the output data at several energies to give a functional form of $1/\omega_0$ and $\kappa$ as functions of $\phi$ and $\varepsilon$.

The Fermi surface for the dispersion Eq.~(\ref{eqn:tightbinding}) 
exhibits a $\phi \rightarrow \phi + \frac{\pi}{2}$ symmetry, 
and so $\kappa(\varepsilon,\phi)$ and $1/\omega_0(\varepsilon,\phi)$ 
also have this symmetry.  This allows us to fit the data to a truncated Fourier series with only $\cos 4n\phi$, $n \in \mathbb{Z}$, terms present.\cite{berg,husseynature}  We find that $1/\omega_0(\varepsilon,\phi)$ may be accurately represented by the form
\begin{eqnarray}
\frac{1}{\omega_0(\varepsilon,\phi)} & = & \frac{1}{\omega_{00}(\varepsilon)} + \frac{1}{\omega_{01}(\varepsilon)}\cos 4\phi + \frac{1}{\omega_{02}(\varepsilon)}\cos 8\phi \nonumber \\
& & +~\frac{1}{\omega_{03}(\varepsilon)}\cos 12\phi
\label{eqn:invomega}
\end{eqnarray}
and $\kappa(\varepsilon,\phi)$ may be parametrized as
\begin{equation}
\kappa(\varepsilon,\phi) = \kappa_0(\varepsilon) + \kappa_1(\varepsilon)\cos 4\phi + \kappa_2(\varepsilon) \cos 8\phi.
\label{eqn:kappa}
\end{equation}
For the energy range we are interested in the energy dependence of each of the coefficients 
can be well approximated as linear, $C(\varepsilon) = C_0 + C_1 \varepsilon$.  
Contributions to the integral [Eq.~(\ref{eqn:result2})] from energies far from the 
Fermi energy are negligible. 
The fitting parameters for $\kappa$ and $1/\omega_0$ are given in Table~\ref{tb:coeffs}.
\begin{table}[htb] 
  \begin{center}
    \caption{Fitting parameters for 
$\kappa(\varepsilon,\phi)$ and $1/\omega_0(\varepsilon,\phi)$.  \label{tb:coeffs}}
    \vspace*{0.2cm}
    \begin{tabular}{ r  r  r  r  r  r  r  r} 
        \hline \hline
         & $1/\omega_{00}$  &$1/\omega_{01}$ & $1/\omega_{02}$& $1/\omega_{03}$& $\kappa_0$ & $\kappa_1$ & $\kappa_2$ \\ \hline
         $C_0$  &   ~4.768 & ~-0.582 & ~0.035 & ~0.035 & ~1.786 & ~-0.077 & ~0.002 \\
         $C_1$  &  ~8.097 & ~2.670 & ~-1.388 & ~-0.361 & ~1.039 & ~0.025 & ~-0.030 \\
        \hline \hline
    \end{tabular}
    \end{center}
\end{table}

The remaining functional inputs we need to calculate the AMRO are the interlayer hopping $t_c(\phi)$ and the scattering rate $1/\tau(\varepsilon,\phi)$.  We assume the standard isotropic scattering rate for $1/\tau(\varepsilon)$:\cite{rammer,agd}
\begin{equation}
\frac{1}{\tau} = \frac{1}{\tau_0} + A\left[(\pi k_B T)^2 + (\varepsilon - \mu)^2\right].
\label{eqn:flscatter}
\end{equation}
We estimate the parameters $1/\tau_0$ and $A$ for $1/\tau$ from fits to the isotropic part of the scattering rate in Ref. \onlinecite{ajawad}, giving $A = (13.5~\mbox{meV})^{-2}$; $1/\tau_0$ always appears as part of a product $1/\omega_{0}(\varepsilon_F)\tau_0$, where $1/\omega_0(\varepsilon_F)$ is the constant contribution to $1/\omega_0(\varepsilon,\phi)$ at the Fermi energy, and we find that  $1/\omega_{0}(\varepsilon_F)\tau_0 = 2.5$.  We assume the hopping is independent of energy and use the same expression for the hopping term as in Ref. \onlinecite{kennettmckenzie}:
\begin{equation}
t_c(\phi) = t_0\left[\sin 2\phi + \eta_1 \sin 6\phi + \eta_2 \sin 10\phi\right].
\label{eqn:hopping}
\end{equation}
The parameters $\eta_1$ and $\eta_2$ are related to each other by the symmetry of the 
crystal lattice, which requires $t_c(\phi) = -t_c(\phi + \frac{\pi}{2})$ and 
hence $\eta_1 = 1 + \eta_2$.  Due to imperfections in the crystal, this relation may not 
hold exactly, and we allow both parameters to vary during fitting.  
The input values are $\eta_1 = 0.737$ and $\eta_2 = -0.263$.  
The constant $t_0$ enters the overall normalization and does not need to be 
specified in our calculation.

\begin{figure}[htb] 
  \includegraphics{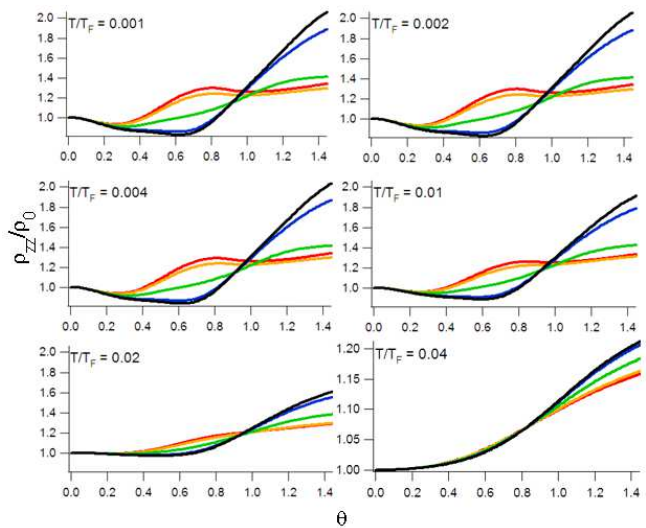}
  \caption{AMRO data calculated numerically using paramters appropriate for thallium cuprate.  The values of $\varphi$ used are 0, 20, 28, 36 and 44 degrees.  Note the change in vertical axis limits in the last plot.  The plots shown correspond to temperatures of 2.8 K, 5.7 K, 11.3 K, 28 K, 56 K and 113 K.}
     \label{fig:fakeamro}
\end{figure}

With explicit functional forms for all functions appearing in
 Eq.~(\ref{eqn:result2}), we may perform the angular and energy integrals.  
Due to the fact that $-\partial f_T/\partial \varepsilon$ decays quickly 
away from $\varepsilon = \mu$ we only integrate over the energy from
 $\mu - 5k_BT$ to $\mu + 5k_BT$ to cover the range of energy that 
contributes to the integral, and we checked that our results were independent 
of the energy range for this choice of integration interval.  
The values of $\varphi$ we use are 0, 20, 28, 36 and 44 degrees 
and we generate AMRO for $T = 0.001\hspace*{0.1cm}T_F - 0.04\hspace*{0.1cm}T_F$, 
where $T_F$ is $\mathcal{O}(3000~\mbox{K})$.  These correspond to 
temperatures from 3 K to 113 K.  
The output is normalized such that $\rho_{zz}(\theta = 0) = 1$.  
The simulated AMRO data are shown in Fig.~\ref{fig:fakeamro}.

\subsection*{Fitting the Simulated Data}

We fit to the simulated AMRO data using Eq.~(\ref{eqn:hopping}) for $t_c(\phi)$ and parameterize the scattering rate as\cite{ajawad,kennettmckenzie}
\begin{equation}
\frac{1}{\tau(\phi)} = \frac{1}{\tau_0}\left[1 + \alpha \cos 4 \phi\right].
\label{eqn:scatterfit}
\end{equation}
The Fermi wavevector and cyclotron frequency are parametrized as
\begin{equation}
k_F(\phi) = k_F\left[1 + k_4 \cos 4 \phi\right],
\label{eqn:wavefit}
\end{equation}
and
\begin{equation}
\frac{1}{\omega_0(\phi)} = \frac{1}{\omega_{00}}\left[1 + u \cos 4 \phi\right].
\label{eqn:omegafit}
\end{equation}
We use the 2.8 K data to fit to the parameters $\alpha$, $k_4$, $u$, $\omega_{00}\tau_0$, $ck_F$, $\eta_1$ and $\eta_2$.  For fits at temperatures above 2.8 K we assume that only the scattering rate is temperature dependent, and fit only to the parameters $\omega_{00}\tau_0$ and $\alpha$.

\begin{figure}[htb] 
  \includegraphics[scale=0.8]{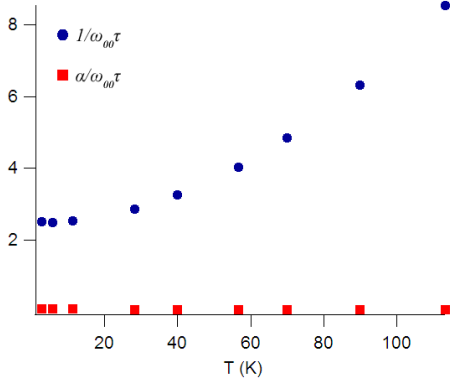}
  \caption{Plot of fit parameters $1/\omega_{00}\tau$ and $\alpha/\omega_{00}\tau$ as functions of temperature.}
     \label{fig:simdatafits}
\end{figure}

The scattering rate parameters $1/\omega_{00}\tau$ and $\alpha/\omega_{00}\tau$ are plotted as a function of temperature in Fig.~\ref{fig:simdatafits}.  A quadratic fit $a(1 + (b\pi k_B T)^2)$ to $1/\omega_{00}\tau$ yields $a = 2.4942$ and $b = 50.88$ K, agreeing with the parameter values used to simulate the AMRO.  The fit values obtained are $k_4 = -0.025$, $u = -0.153$, $ck_F = 8.607$, $\eta_1 = 0.764$ and $\eta_2 = -0.245$, which are within $1-2\%$ of the input parameter values used in the simulation.

We find that not only do the fits to the simulated data fail to reproduce the
 $T\cos 4\phi$ term observed in the experiments, we observe essentially no temperature dependence 
of the anisotropy parameter $\alpha$, suggesting that if the scattering rate were in fact isotropic the fitting procedure used by Abdel-Jawad \emph{et al.}\cite{ajawad} would not have yielded anisotropic scattering.  Hence, the anisotropic contribution to the electron-electron scattering rate measured in recent AMRO experiments is not an artifact of the fitting procedure, and the system does deviate from standard 
Fermi liquid behavior.

\section{Conclusion}
\label{sec:conclusion}

In this work we present expressions for the interlayer conductivity of layered metals 
with either an isotropic or anisotropic Fermi surface, allowing for thermal effects.
We used the expression for the conductivity for a layered metal with 
anisotropic Fermi surface properties to simulate the AMRO expected an isotropic 
scattering rate consistent with Fermi liquid theory.  Fitting to the simulated 
data using the same procedure used to fit the experimental data in 
Ref.~\onlinecite{ajawad} does not reproduce the anisotropic $T\cos 4\phi$ piece 
of the scattering rate.  Hence, we have confirmed that including finite 
temperature effects whilst taking into account anisotropy in the 
dispersion does not affect the conclusion there there is an 
anisotropic contribution to the scattering rate of thallium cuprate.

The theory presented here is applicable to any layered metal in which 
AMRO may be observed, and provides the means to analyze AMRO data for temperatures at which 
$T/T_F$ grows large enough that finite temperature effects on AMRO become 
 significant.  This may allow for a more precise interpretation 
of current AMRO data at higher temperatures, for example, in cuprate experiments\cite{private}
 where the maximum value of 
$T/T_F$ is $\sim0.03-0.04$ or in Na$_{0.48}$CoO$_2$, 
which has a Fermi energy around 250 meV,\cite{qian} and AMRO experiments
have been performed at temperatures as high as $T/T_F \sim 0.02$.\cite{huetal} 
This theory will also enable accurate quantitative analysis of future high temperature AMRO experiments.

\section*{Acknowledgments}
We thank R. McKenzie and N. Hussey for helpful discussions on AMRO.  
This work was supported by NSERC.


\begin{thebibliography}{99}
\bibitem{kart} M. V. Kartsovnik, Chem. Rev. \textbf{104}, 5737 (2004).

\bibitem{berg} C. Bergemann, A. P. Mackenzie, S. R. Julian, D. Forsythe 
and E. Ohmichi, Adv. Phys. \textbf{52}, 639 (2003).

\bibitem{balicas} L. Balicas, S. Nakatsuji, D. Hall, T. Ohnishi, Z. Fisk, Y. Maeno and D. J. Singh, Phys. Rev. Lett. \textbf{95}, 196407 (2005).

\bibitem{osada} T. Osada, H. Nose and M. Kuraguchi, Physica B \textbf{294}-\textbf{295}, 402 (2001).

\bibitem{beierl} U. Beierlein, C. Shenkler, J. Dumans and M. Greenblatt, Phys. Rev. B \textbf{67}, 235110 (2003).

\bibitem{husseynature} N. E. Hussey, M. Abdel-Jawad, A. Carrington, A. P. Mackenzie and L. Balicas, Nature (London) \textbf{425}, 814 (2003).

\bibitem{enomoto} K. Enomoto, S. Uji, T. Yamaguchi, T. Terashima, T. Konoike, M. Nishimura, T. Enoki, M. Suzuki and I. S. Suzuki, Phys. Rev. B \textbf{73}, 045115 (2006).  

\bibitem{Sandemann} K. G. Sandemann and A. J. Schofield, Phys. Rev. B {\bf 63}, 094510 (2001).

\bibitem{ajawad} M. Abdel-Jawad, M. P. Kennett, L. Balicas, A. Carrington, A. P. Mackenzie, R. H. McKenzie and N. E. Hussey, Nature Physics \textbf{2}, 821 (2006).

\bibitem{kennettmckenzie} M. P. Kennett and R. H. McKenzie, Phys. Rev. B, \textbf{76}, 054515 (2007).
\bibitem{ajawad2} M. Abdel-Jawad, J. G. Analytis, L. Balicas, J. P. H. Charmant, M. M. J. French and N. E. Hussey, Phys. Rev. Lett. \textbf{99}, 107002 (2007).

\bibitem{analytisetal} J. G. Analytis, M. Abdel-Jawad, L. Balicas, M. M. J. French and N. E. Hussey, Phys. Rev. B, \textbf{76}, 104523 (2007).

\bibitem{kennett2} M. P. Kennett and R. H. McKenzie, Physica B, \textbf{403}, 1552 (2008).

\bibitem{hussey2} N. E. Hussey, M. Abdel-Jawad, L. Balicas, M. P. Kennett and R. H. McKenzie, Physica B \textbf{403}, 982 (2008). 

\bibitem{singleton} J. Singleton, P. A. Goddard, A. Ardavan, A. I. Coldea, S. J. Blundell, R. D. McDonald, S. Tozer and J. A. Schlueter, Phys. Rev. Lett. \textbf{99}, 027004 (2007).

\bibitem{smithmckenzie} M. F. Smith and R. H. McKenzie, Phys. Rev. B {\bf 77}, 235123 (2008).

\bibitem{kondo} T. Kondo, T. Takeuchi, S. Tsuda and S. Shin, Phys. Rev. B \textbf{74}, 224511 (2006).

\bibitem{plate} M. Plat\'{e}, J. D. F. Mottershead, I. S. Elfimov, D. C. Peets, R. Liang, 
D. A. Bonn, W. N. Hardy, S. Chiuzbaian, M. Falub, M. Shi, L. Patthey, and A. Damascelli,
  Phys. Rev. Lett. \textbf{95}, 077001 (2005); D. C. Peets, J. P. F. Mottershead, B. Wu, I. S. Elfimov, R. Liang, W. N. Hardy, D. A. Bonn, M. Raudsepp, N. J. C. Ingle and A. Damascelli, New. J. Phys \textbf{9}, 28 (2007).

\bibitem{kaminski} A. Kaminski,
 H. M. Fretwell, M. R. Norman, M. Randeria, S. Rosenkranz, U. Chatterjee, 
J. C. Campuzano, J. Mesot, T. Sato, T. Takahashi, T. Terashima, M. Takano, K. Kadowaki, 
Z. Z. Li, and H. Raffy, Phys. Rev. B {\bf 71}, 014517 (2005).

\bibitem{Chang} J. Chang, 
M. Shi, S. Pailh\'{e}s, M. Mansson, T. Claesson, O. Tjernberg, A. Bendounan, Y. Sassa, L. Patthey,
N. Momono, M. Oda, M. Ido, S. Guerrero, C. Mudry, and J. Mesot, Phys. Rev. B {\bf 78}, 205103 (2008).

\bibitem{hac} N. E. Hussey, J. C. Alexander and R. A. Cooper, Phys. Rev. B \textbf{74}, 214508 (2006).

\bibitem{narduzzo} A. Narduzzo, G. Albert, M. M. J. French, N. Mangkorntong, M. Nohara, H. Takagi and N. E. Hussey, Phys. Rev. B {\bf 77}, 220502(R) (2008).

\bibitem{ioffe} L. B. Ioffe, A. J. Millis, Phys. Rev. B, \textbf{58}, 11361 (1998).

\bibitem{civelli} M. Civelli, M. Capone, S. S. Kancharla, O. Parcollet and G. Kotliar, Phys. Rev. Lett. \textbf{95}, 106402 (2005). 

\bibitem{private} N. E. Hussey, private communication.

\bibitem{footnote} For the purposes of illustration we ignore possible energy 
dependence of the interlayer hopping $t_c$ for an isotropic Fermi surface, but
we do include this possibility in our more general analysis of a layered metal
with anisotropic Fermi surface properties.

\bibitem{mckenziemoses}  R. H. McKenzie and P. Moses, Phys. Rev. Lett. \textbf{81}, 4492 (1998).

\bibitem{yagi} R. Yagi, Y. Iye, T. Osada, S. Kagoshima, J. Phys. Soc. Jpn., \textbf{59}, 3069 (1990).

\bibitem{ashcroftmermin} N. W. Ashcroft and N. D. Mermin, ``Solid State Physics'', Harcourt College Publishers Inc. (1976).

\bibitem{mosesandmckenzie} P. Moses and R. H. McKenzie, Phys. Rev. B \textbf{60},  7998 (1999).

\bibitem{Yamaji} K. Yamaji, J. Phys. Soc. Jpn. {\bf 58}, 1520 (1989).

\bibitem{kennett4} M. P. Kennett and R. H. McKenzie, Phys. Rev. B {\bf 78}, 024506 (2008).

\bibitem{rammer} J. Rammer, \emph{Quantum Transport Theory}, Westview Press (2004).

\bibitem{agd} A. A. Abrikosov, L. P. Gorkov and I. E. Dzyaloshinski, \emph{Methods of Quantum Field Theory in Statistical Physics}, Dover Publications, Inc., New York.  (1963).

\bibitem{qian} D. Qian, L. Wray, D. Hsieh, L. Viciu, R. J. Cava, J. L. Luo, D. Wu, N. L. Wang, and 
M. Z. Hasan, Phys. Rev. B \textbf{97}, 186405 (2006).

\bibitem{huetal} F. Hu, G. T. Liu, J. L. Luo, D. Wu, N. L. Wang, and T. Xiang, Phys. Rev. B \textbf{73}, 212414 (2006).





\end{thebibliography}
\end{document}